\documentclass[preprint,showpacs,preprintnumbers,amsmath,amssymb]{revtex4}

\usepackage{graphicx}
\usepackage{dcolumn}
\usepackage{bm}

\begin{document}

\begin{titlepage}

\title{Time dependent correlations in a supercooled
liquid from nonlinear fluctuating hydrodynamics.}
\thispagestyle{empty}

\author{Bhaskar Sen Gupta$^1$, Shankar P. Das$^1$ and Jean-Louis Barrat$^2$}

\affiliation{$^1$School of Physical Sciences, Jawaharlal Nehru University, New Delhi - 110067, India \\
$^2$ Universit\'e de Lyon; Univ. Lyon I,  Laboratoire de
Physique de la Mati\`ere Condens\'ee et des Nanostructures; CNRS,
UMR 5586, 69622 Villeurbanne Cedex,
France.}

\begin{abstract}
We solve numerically the equations of nonlinear fluctuating
hydrodynamics (NFH) for the supercooled liquid. The time correlation
of the density fluctuations in equilibrium obtained here shows
quantitative agreement with molecular dynamics(MD) simulation data.
We demonstrate numerically that the $1/\rho$ nonlinearity in the NFH
equations of motion is essential in restoring the ergodic behavior
in the liquid. Under nonequilibrium conditions the time correlation
functions  relax in a manner similar to that observed in the
molecular dynamics simulations in binary mixtures. The waiting time
$t_\mathrm{w}$ dependence of the non-equilibrium response function
follows a Modified Kohlrausch-Williams-Watts(MKWW) form similar to
the behavior seen in dielectric relaxation data.
\end{abstract}


\maketitle
\end{titlepage}

The conserved densities of mass, momentum, and energy
constitute the simplest set of slow modes characteristic of the
isotropic liquid. The microscopic balance equations for the
respective conservation laws contain terms with widely different
characteristic time scales of variation respectively corresponding
to the various degrees of freedom of the complex system.  The
nonlinear fluctuating hydrodynamics (NFH) describes the dynamics of
these slow modes with nonlinear differential equations having
regular and stochastic parts. The regular parts involve nonlinear
coupling of slow modes while the random parts represent noise which
can be linear\cite{DM,ABL,hayakawa} or
multiplicative\cite{ABL,kim-kawasaki}. The most widely studied
theoretical model for the slow dynamics in a supercooled liquid
approaching vitrification follows from these equations of NFH and is
termed as the self-consistent mode coupling theory (MCT)
\cite{rmp,reichman}. In a strongly interacting dense liquid the
coupling of density fluctuations produces the dominant effect on
dynamics. The MCT involves a nonlinear feedback mechanism\cite{beng}
of density fluctuations  producing strong enhancement of the
viscosity of the supercooled liquid. In its simplest version the MCT
predicts that above a critical density the long time limit of the
time correlation ${\cal C}(t)$ of density fluctuations is nonzero.
This signifies an ergodic-nonergodic transition(ENE) in the liquid
and is a precursor to the liquid-glass transition. The predicted
dynamics involves several different regimes of relaxation and has
been widely used in fitting experimental data on different liquids.
However, the simple MCT approach is known to exaggerate the tendency
of the dynamics towards slowing down, and becomes quantitatively
inaccurate in the vicinity of the predicted transition, which is
never observed in practice. The perturbation expansion for the
renormalized transport coefficient in the MCT, though systematic, is
in terms of a dimensionless parameter which is not small.
Furthermore, it has also been shown\cite{DM,DM1} non perturbatively
that the $1/\rho$ nonlinearities in the NFH equations remove the
sharp ENE transition predicted in the simplified theory. We report
here the study of the slow dynamics of a dense monatomic
Lennard-Jones liquid by numerically solving the stochastic equations
of NFH. Our nonperturbative calculation shows good agreement with
the computer simulation results of the same system in equilibrium.
We also report here results on the dynamics of the density
fluctuations under nonequilibrium conditions.

We consider the equations of NFH for an isotropic liquid in its
simplest form for the mass density $\rho$ and momentum density
\textbf{g}\cite{DM}

\begin{eqnarray}
\label{cont} &&\frac{\partial\rho}{\partial{t}} + {\bf \nabla}.{\bf
g} = 0, \\
\label{g_eq} && \frac{\partial g_{i}}{\partial t} + \nabla_j \left [
\frac{g_ig_j}{\rho} \right ] + \rho \nabla_{i} \frac{\delta
F_U}{\delta \rho} + L_{ij} \frac{g_j}{\rho} = \theta_{i}~~.
\end{eqnarray}

\noindent  The correlations of the gaussian noise $\theta_{i}$ are
related to the bare damping matrix $L_{ij}$ \cite{hansen},
$\left\langle \theta_{i}(x,t)\theta_j
({x^\prime}t^\prime)\right\rangle = 2k_{B}TL_{ij}
\delta(t-t^\prime)\delta(x-x^\prime)$. For an isotropic liquid,
$L_{ij} = (\zeta_{0} + \eta_{0}/3)\delta_{ij}\nabla^{2} +
\eta_{0}\nabla_{i}\nabla_{j}$ where  $\zeta_{0}$ and $\eta_0$
respectively denote is the bare bulk and shear viscosities. The
stationary solution to the Fokker-Planck equation corresponding to
the generalized Langevin eqn. (\ref{g_eq}) is obtained as
$\exp\{-\beta{F[\rho,g]}\}$ with $\beta=1/{k_{B}T}$ is the Boltzmann
factor. The coarse grained free energy functional is obtained as
$F[\rho,g]\equiv F_K[\rho,g]+F_U$. The kinetic part is dependent on
the momentum density $F_K=\int d{\bf x} g^2/(2\rho)$ and the so
called potential part is given by
$F_U=F_\mathrm{id}+F_\mathrm{int}$. The ideal gas contribution is
$F_\mathrm{id} = \int d{\bf r} \rho({\bf r}) [ \ln ( \rho({\bf
r})/\rho_{0}) - 1 ]$. The interaction part $F_\mathrm{int}$  up to
quadratic order in density fluctuations\cite{tvr} is obtained as
\begin{equation}
 \beta F_\mathrm{int} = - \frac{1}{2m^2} \int
d{\bf r} d {\bf
r}^{'}c(\textbf{r}-\textbf{r}^{'})\delta\rho(\textbf{r})
\delta\rho(\textbf{r}^{'})~~,\label{R_Y}
\end{equation}

\noindent  where $c(r)$ is the two point Ornstein-Zernike direct
correlation function\cite{hansen} and $m$ is the mass of the
particles. For the glassy dynamics we focus on the coupling of
slowly decaying density fluctuations present in the pressure
functional, represented by the third term on the LHS of eqn.
(\ref{g_eq}). With the above choice of $F_U$, the nonlinear
contribution in this term reduces to $\rho\nabla_i f(r,t)$ with the
convolution $f(\textbf{r},t) = m^{-1}\int d\textbf{r}
c(\textbf{r}-\textbf{r}^{'}) \delta \rho(\textbf{r}^{'},t)$.

We consider here a classical system of $N$ particles, each of mass
$m$ interacting with a Lennard-Jones potential $u(r) = 4\epsilon [
{(\sigma/r)}^{12}- {(\sigma/r)}^{6} ]$.  In addition to the scale
$\sigma$ of the interacting potential there is another length $h$ of
the lattice grid on which $\rho$ and ${\bf g}$ are computed. We
choose $\sigma/h$ to be non integer ( $=4.6$ in the present
calculation) to avoid crystallization. Time is scaled with the LJ
unit of $\tau_{0} = (m\sigma^{2}/\epsilon)^{\frac{1}{2}}$ and length
with $h$. The thermodynamic state of the fluid is described in terms
of the reduced density $n^{*} = n_{0}\sigma^{3}$ and the reduced
$T^{*} = (k_{B}T)/\epsilon$. For numerical solution the conserved
densities are scaled to dimensionless forms: $n(\textbf{r}) = [
h^3m^{-1}] \rho(\textbf{r})$,  and $\textbf{j}(\textbf{r}) = [
h^{3}(m\epsilon)^{-\frac{1}{2}}] \textbf{g}(\textbf{r})$.  The speed
of sound $c_0$ is given by, $c_0^2=k_{B}T/(mS(0))$. We start with an
initial distribution of the fluctuating variables $n(\textbf{r})$
and $\textbf{j}(\textbf{r})$ over a set of points $20^3$ on a cubic
lattice. The nonlocal integral $f(r,t)$ is evaluated as a sum of
contributions from the successive shells, $ f(r,t) = h^{3}\sum_{i}
c(R_{i})\sum_{\alpha}\delta n(R_{i}^{\alpha},t)$, where
$R_{i}^{\alpha}$ for $\alpha=1,...m_i$ respectively denote radii
vectors of the $m_i$ lattice points in the \textit{i}th spherical
shell of radius $R_i$. The $1/\rho$ nonlinearity in the dissipative
term of the momentum equation is computed by replacing the density
field in the denominator with the $\rho({\bf x})$ averaged over a
length scale close to $\sigma$ around the corresponding point ${\bf
r}$. We ignore the convective nonlinearity in the present
calculation and focus on the role of the pressure nonlinearity in
producing the slow dynamics.

A major hurdle encountered in the numerical scheme used here arises
from an instability which occurs as $n({\bf x},t)$ gets negative at
certain grid points. To avoid this situation, we redefine $n({\bf
x},t)$ on the grid at each step of the numerical integration with a
coarse graining scheme. In devising the latter we make use of the
following physical interpretation of the definition of $\rho({\bf
x},t)$ of the density field : the integral $\int_{\Delta{V}} d{\bf
x} \rho({\bf x},t)$ represents the total mass in an elementary
volume $\Delta{V}$ of the system. At each time step of the numerical
integration, the positivity of the field $n({\bf x})$ over the whole
grid is checked. If it turns negative at a point, we reduce $n({\bf
x})$ at some or all of the neighboring sites by taking equal
contributions from each and add the sum total to the original site.
It is also ensured that the density at none of the neighboring sites
becomes negative as a result of this redistribution. The sum of the
densities at the original and the contributing sites remains
unaltered and hence global conservation is maintained. If the above
redistribution involving contributions only from the nearest
neighbor sites is insufficient to make $n({\bf x})$ positive
everywhere, we include the next nearest neighbors in the
redistribution and so on.  In reality however we hardly need to
include beyond the second shell of neighbors surrounding the
original site. With the density instability being corrected with
this coarse graining procedure, the numerical algorithm can be  run
up to much larger times  than  in earlier works\cite{dasgupta}. The
arbitrary regularization of  the strength of the noise
\cite{dasgupta} can also be avoided,  and the fluctuation
dissipation relation respected.

The equal-time correlation of density fluctuations for the $N$
particle system is given by $S(k,t) = {N}^{-1} < \delta n(k,t)
\delta n(-k,t)>$ where the angular brackets refer to an average over
the noise. We first consider the system as it evolves at $T^*=2.0$
and $n_0^*=.97$ under the influence of thermal noise, starting from
an initial state in which all fluctuations are set to zero. Time
translational invariance is reached as the system equilibrates and
$S(k,t)$ approaches the corresponding static structure factor
$S(k)$. The equilibrated $S(k)$ vs. $k$ plot ( for large
$t_\mathrm{w}$ ) is displayed in fig. 1. $S(k,t)$ obtained
with equations of motion linear in fluctuations is also displayed.
The peak position ($q_m$) and amplitude of the $S(k)$ obtained (
using the Ornstein-Zernike relation) from the input direct
correlation function $c(r)$ are well reproduced. Other features of
$S(k)$ are partly lost due to the relatively crude grid size used in
our numerical solution.

\begin{figure}
\begin{center}
\includegraphics*[width=8cm]{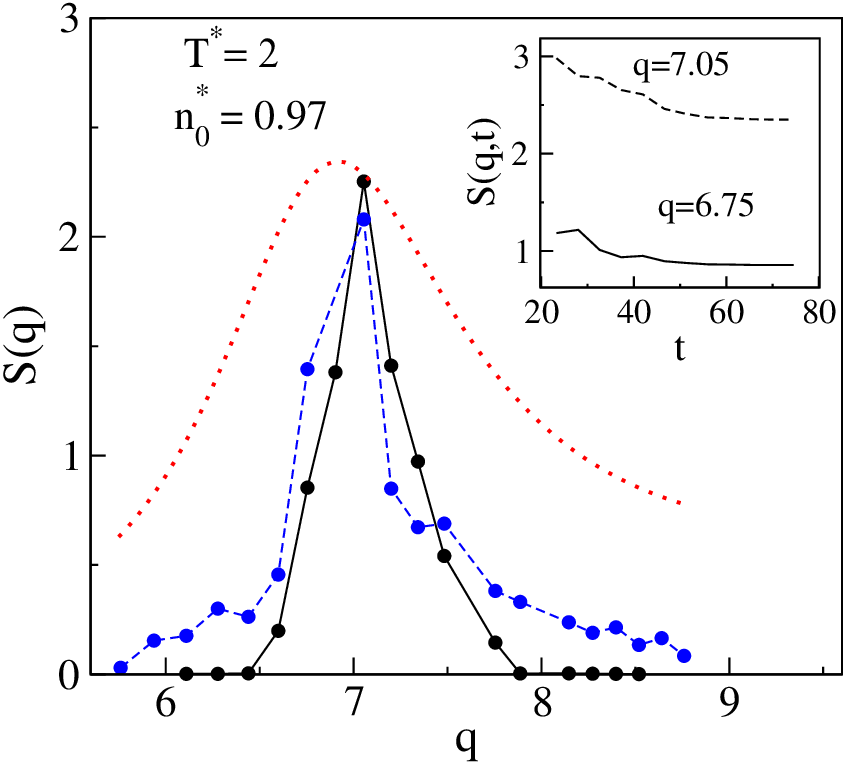}
\caption{ $S(k)$ vs $k\sigma$ at $T^*=2$ and $n_0^*=0.97$ for linear
(solid) and nonlinear (dashed) dynamics; as computed from input
$c(r)$ (dotted). Inset: $S(k,t)$ vs $t$ for $k\sigma=6.75$ (solid)
and $k=7.05$ (dashed) displaying equilibration with time.}
\label{fig1}
\end{center}
\end{figure}

\noindent Next we focus on the dynamic correlation  function
${C}(t+t_\mathrm{w},t_\mathrm{w})$ defined in the normalized form
${C}(t+t_\mathrm{w},t_\mathrm{w})= {<\delta n(t+t_\mathrm{w}) \delta
n(t_\mathrm{w})>}/{<\delta n(t_\mathrm{w}) \delta
n(t_\mathrm{w})>}.$
For large $t_\mathrm{w}$ time translational invariance holds, making
$ {C}(t+t_\mathrm{w},t_\mathrm{w}) \equiv {\cal C}(t)$. The decay of
${\cal C}(q_m,t)$ is compared in fig. 2 with the corresponding
molecular dynamics simulation results\cite{ullo} for the
equilibrated systems at $T=.6$ for two densities $n_0^*=1.10$ and
$n_0^*=1.06$. The input $c(r)$ corresponds to the purely repulsive
part of the Lennard-Jones potential following Ref. \cite{ullo}. The
bare transport coefficients which determine the noise correlations
are chosen such that the short time dynamics agrees with computer
simulation results.  ${\cal C} (t)$ obtained by solving the
stochastic equations linearized in the fluctuations decays very
fast. In comparison considerable slowing down of the decay of ${\cal
C}(t)$ occurs on solving the full NFH equations. The static
correlation function $S(k)$ however ( see inset of fig. 3) shows
hardly any difference between the two cases. At higher densities the
mean-free path of the fluid particles gets smaller and approaches
the atomic length scale. As a result, the validity of Generalized
hydrodynamic equations at short length scales ( corresponding to
wave vector $q{\sim}q_m$ ) improves with increasing density. This
trend is clearly seen our results displayed in fig. 2 for ${\cal
C}(t)$.

The ENE transition of simple MCT  is driven by the nonlinear
couplings of density fluctuations in the pressure term  (3rd term on
LHS) of the generalized Navier-Stokes equation (\ref{g_eq}). On the
other hand the $1/\rho$ nonlinearity crucial for the absence of the
ENE transition is in the dissipative term of the same equation, {\em
i.e.}, 4th term on LHS of (\ref{g_eq}). We therefore consider two
cases here to test the role of the relevant nonlinearities from a
non perturbative approach. In Case A the $1/\rho$ nonlinearity in
(\ref{g_eq}) is replaced with $1/\rho_0$ while keeping the density
nonlinearity in the pressure term. In Case B, the complete model
with both nonlinearities is considered. The results for ${\cal
C}(q_m, t)$  at $T=.6$ and $n_0^*=1.10$ corresponding to the Cases A
and B respectively are shown in fig 3. We extend the numerical
solution to the longest possible time scale ( $>10^3$ in
Lennard-Jones units) which is about four orders of magnitude beyond
the microscopic time scales. The decay of the dynamic correlation is
markedly different in the two cases and agrees with the previous
theoretical results on the role of $1/\rho$ nonlinearity.

To  study the structural relaxation in the nonequilibrium state we
consider the time evolution of the Lennard-Jones liquid following an
instantaneous quench from $T^*_{i}=2.0$  and $n_0^*=0.97$ along the
isobaric line to $T^*_{f}=0.4$ and $n_0^*=1.12$. We compute from the
solution of the NFH equations the two-time density correlation
function $C(t_\mathrm{w}+t,t_\mathrm{w})$ at $q=q_m$ for different
waiting times $t_\mathrm{w}=50,$ $100,$ $200,$ $500,$ $700$ and
$1000$. For small values of $t$ time translational invariance holds
and $C(t_\mathrm{w}+t,t_\mathrm{w})$ depends only on $t$. On the
other hand at large $t$, the correlation function depends on both
$t$ and $t_\mathrm{w}$. Following the mean-field-theoretical
results\cite{spin} and also experimental data\cite{expt} fits on
spin glasses, we fit this long time part of the density correlation
function with the form $ C^{ag}\left[
{h(t+t_\mathrm{w})}/{h(t_\mathrm{w})}\right]$, where $h(t)$ is a
monotonously ascending function of its argument. In the
$C(t_\mathrm{w}+t,t_\mathrm{w})$ vs.
$[h(t+t_\mathrm{w})/h(t_\mathrm{w})]^\alpha$ plot, the parameter
$\alpha$ is tuned to obtain a collapse of all the curves. The
results displayed in the fig. 4 obtain  $h(t) \sim
[\rm{log}(t)]^\alpha$ with the best fit value of the parameter
$\alpha=.81$. This is comparable  to the corresponding value
$\alpha=.88$ obtained in molecular dynamics simulations \cite{KB} of
binary Lennard-Jones mixtures.

In a recent work, Lunkenheimer et. al. \cite{loidl} studied over a
range of frequency $\omega$ the dielectric response function
$\chi_\omega(t_\mathrm{w})$ at temperature $T<T_g$. The system falls
out of equilibrium over laboratory time scales at the calorimetric
glass transition temperature $T_g$. The aging time
($\equiv{t_\mathrm{w}}$ in the present notation) dependence of
$\chi$ follows a modified Kohlrausch-Williams-Watts (MKWW) function
$\tilde{f}(t_\mathrm{w}) =
\exp[{(t_\mathrm{w}/\tau(t_\mathrm{w}))}^{\bar{\beta}}]$. The
relaxation time $\tau(t_\mathrm{w})$ and the stretching exponent
$\bar{\beta}$ are identical  for all frequencies. The limiting value
$\tau(t_\mathrm{w}\rightarrow\infty)$ is close to the
$\alpha$-relaxation time $\tau_\alpha$ extrapolated to corresponding
temperature $T<T_g$\cite{loidl}. In the present work $T\sim{T_c}$
($>T_g$) and in this case the system in fact equilibrates. We study
here the function $\chi_\omega(t_\mathrm{w})=\omega
C(\omega,t_\mathrm{w})$ which in equilibrium would reduce to the
corresponding response function. $C(\omega,t_\mathrm{w})$ is
obtained approximately (i.e. ignoring FDT violations) from the
frequency transform of $C(t+t_\mathrm{w},t_\mathrm{w})$ with respect
to $t$. The data is fitted to the form :
\begin{equation} \label{fit-fun}
\chi_\omega(t_\mathrm{w})= [\chi_\omega^{st} - \chi_\omega^{eq} ]
\tilde{f}(t_\mathrm{w})+ \chi_\omega^{eq},
\end{equation} \noindent
where $\chi_\omega^{st}$ and $\chi_\omega^{eq}$ respectively refer
to the initial and final values of $\chi_\omega$. For the relaxation
time in $\tilde{f}$ we use\cite{sen_das} $\tau(t_\mathrm{w}) =
\left( \tau_{\mathrm{st}}- \tau_{\mathrm{eq}}\right) f(t_\mathrm{w})
+ \tau_{\mathrm{eq}}$ with the normalized function $f(t)=
2^{\bar{\beta}}/{[1+\exp\{2t/\tau(t)\}]}^{\bar{\beta}}$.
Fig. 5 shows that $\chi_\omega$'s at different frequencies
scale onto a single master curve  for $\beta= 0.68$. The
$\tau(t_\mathrm{w})$ decreases sharply with $t_\mathrm{w}$ initially
and becomes almost constant at $\tau_\mathrm{eq}$. At large
$t_\mathrm{w}$ the relaxation  follows a stretched exponential form
having the $\alpha$-relaxation time $\tau_\mathrm{eq}$ and exponent
$\bar{\beta}$ at the corresponding temperature. This shown in the
inset of fig. 5.

\begin{figure}[!ht]
\begin{center}
\includegraphics*[width=8cm]{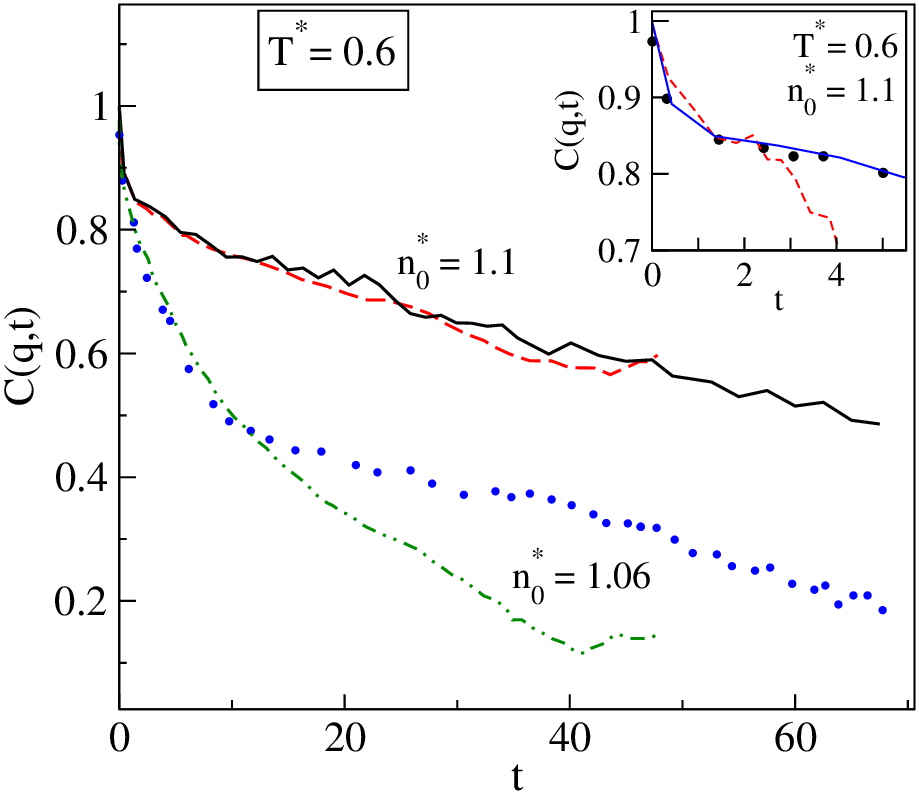}
\caption{  ${\cal C}(q,t)$ vs $t/\tau_0$ at $T^*=.6$  for densities
$n_0^*=1.10$ (solid) and $1.06$ (dotted). Corresponding MD
simulation data\cite{ullo} shown respectively with (dashed) and
(dot-dashed) curves. Inset : short time part of ${\cal C}(q,t)$ with
(solid) and without (dashed) nonlinear coupling of modes;
corresponding MD simulation result\cite{ullo} (dark circles). }
\label{fig2}
\end{center}
\end{figure}

\begin{figure}[!ht]
\begin{center}
\includegraphics*[width=8cm]{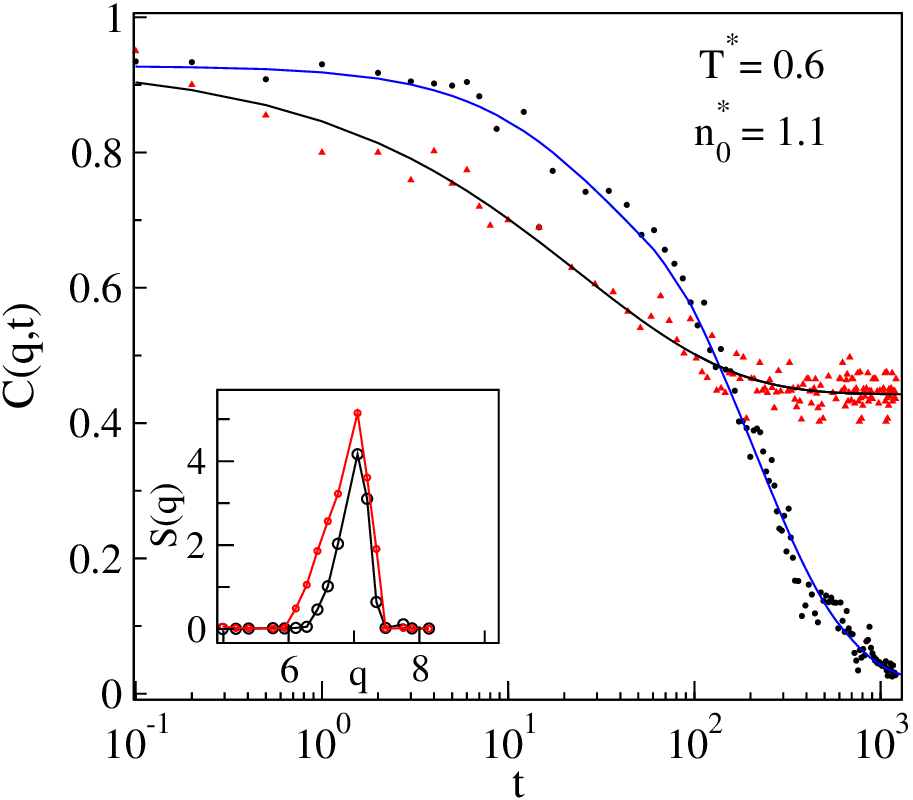}
\caption{ ${\cal C}(q,t)$ vs $t/\tau_0$ at $T^*=.6$ and $n_0^*=1.10$
for case A (filled circles) and case B (filled triangles). The solid
lines are the best fit curves to the corresponding data. Inset :
$S(k)$ vs $k\sigma$ for the two cases respectively with solid and
dashed curves. } \end{center} \label{fig3}
\end{figure}

\begin{figure}[!ht]
\begin{center}
\includegraphics*[width=8cm]{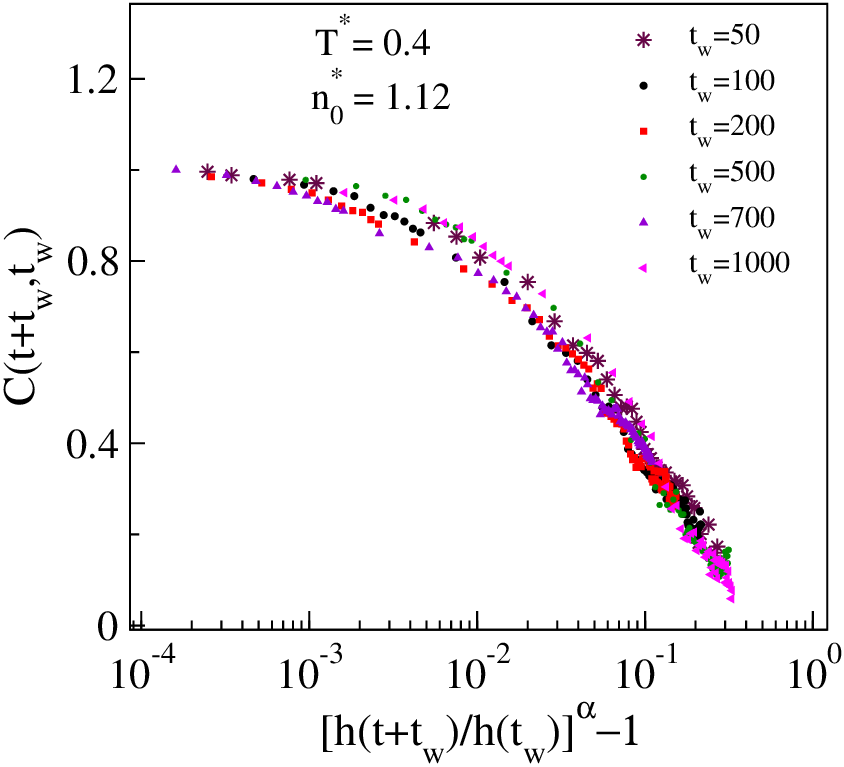}
\caption{Normalized correlation function
$C(t_\mathrm{w}+t,t_\mathrm{w})$ vs.
${[h(t+t_\mathrm{w})/h(t_\mathrm{w})]}^\alpha$ for different waiting
times $t_\mathrm{w}/\tau_0$ shown in the inset . The data at large
$t$ scales onto a single master curve with $h(t)\sim\log(t)$ and
$\alpha=.81$.}
\end{center}
\label{fig4}
\end{figure}

\begin{figure}[!ht]
\begin{center}
\includegraphics*[width=8cm]{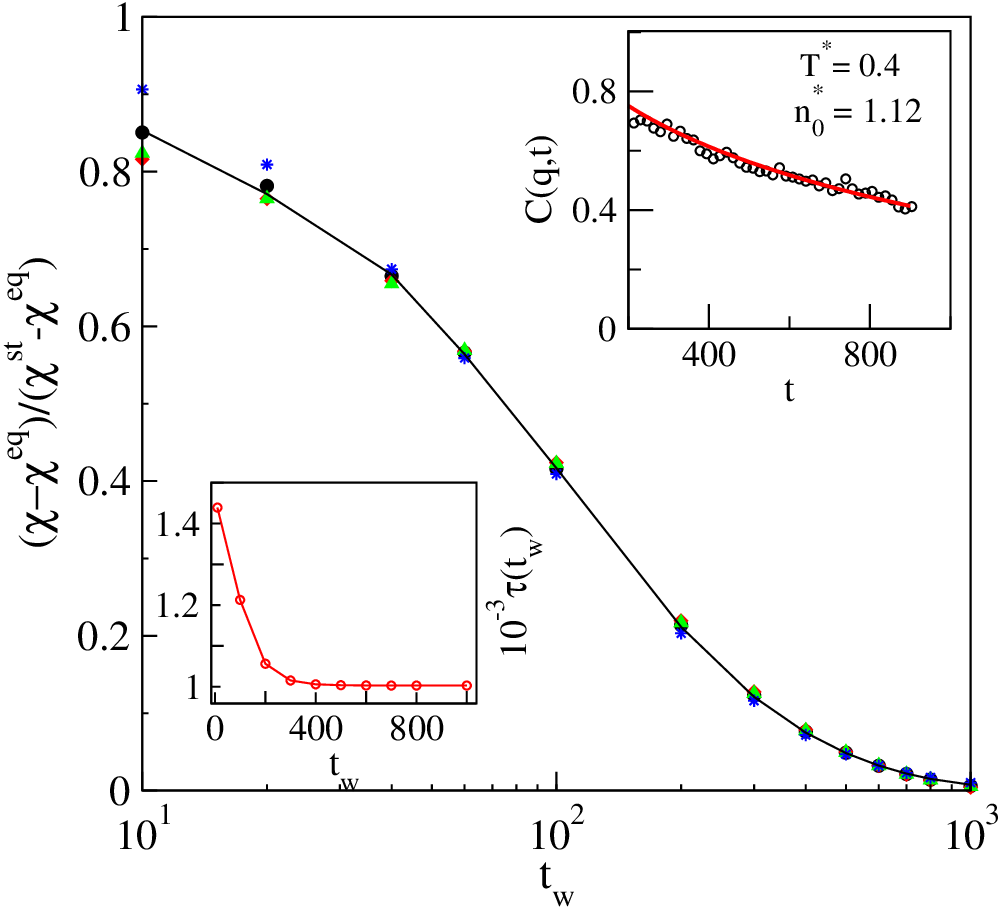}
\caption{Scaling of $\chi_\omega(t_\mathrm{w})$  for frequencies
$\omega\tau_0$ = $10^{-2}$(circle), $10^{-1}$(diamond), $10^0$
(triangle), and $10^1$ (star). solid line is a fit to MKWW form.
Lower inset : $\tau(t_\mathrm{w})$ vs $t_\mathrm{w}/\tau_0$. Upper
inset : ${\cal C}(q_m,t)$ vs $t/\tau_0$ at the final state $T^*=0.4$
and $n_0^*=1.10$ (circles); solid line is fit to a KWW form with
exponent $\bar{\beta}=0.68$ and relaxation time $\tau_{eq}$(see
text).}
\end{center}
\label{fig5}
\end{figure}

We have shown that the direct numerical solutions of the NFH
equations provide a reliable way of studying the dynamics of
fluctuations in a dense liquid in the vicinity of the avoided
ergodic nonergodic transition. While the method is not appreciably
more efficient than molecular dynamics from a computational
standpoint, it provides an interesting way of investigating
theoretical assumptions that can be made in the analytical treatment
of these equations. In particular, our work clearly shows the role
of the $1/\rho$ nonlinearity present in the dissipative term of eqn.
(\ref{g_eq}) in restoring ergodicity \cite{DM}. The present method
can be easily extended to a larger set of hydrodynamic variables,
which would permit a description of the dynamics in binary mixtures.
For the nonequilibrium states, the dynamics with the one loop mode
coupling theory has been formulated for the spherical p-spin glass
model\cite{kurchan}. For the supercooled liquid a similar analysis
even at the one loop level is still lacking. Our numerical approach
is by essence non perturbative, and it will be interesting to
compare it to analytical perturbative approaches to the same
equations under nonequilibrium conditions. The CEFIPRA is
acknowledged for financial support under Indo-French research
project 2604-2.

\end{document}